\documentclass{svproc}
\usepackage{url}
\usepackage{graphicx}
\usepackage{tabularx}

\begin{document}
\mainmatter              
\title{Real-Time Multimodal Data Collection Using Smartwatches and Its Visualization in Education}
\titlerunning{Real-Time Multimodal Data Collection}  
%
\author{Alvaro Becerra\inst{1} \and Pablo Villegas\inst{1} \and Ruth Cobos\inst{1}}
\authorrunning{Becerra et al.} 
%
%
\institute{Universidad Autónoma de Madrid, Spain \email{\{alvaro.becerra, pablo.villegas,ruth.cobos\}@uam.es}}

\maketitle              

\begin{abstract}
Wearable sensors, such as smartwatches, have become increasingly prevalent across domains like healthcare, sports, and education, enabling continuous monitoring of physiological and behavioral data. In the context of education, these technologies offer new opportunities to study cognitive and affective processes such as engagement, attention, and performance. However, the lack of scalable, synchronized, and high-resolution tools for multimodal data acquisition continues to be a significant barrier to the widespread adoption of Multimodal Learning Analytics in real-world educational settings. This paper presents two complementary tools developed to address these challenges: Watch-DMLT, a data acquisition application for Fitbit Sense 2 smartwatches that enables real-time, multi-user monitoring of physiological and motion signals; and ViSeDOPS, a dashboard-based visualization system for analyzing synchronized multimodal data collected during oral presentations. We report on a classroom deployment involving 65 students and up to 16 smartwatches, where data streams including heart rate, motion, gaze, video, and contextual annotations were captured and analyzed. Results demonstrate the feasibility and utility of the proposed system for supporting fine-grained, scalable, and interpretable Multimodal Learning Analytics in real learning environments.

\end{abstract}
\section{Introduction}

The rapid proliferation of wireless sensor technologies in recent years has enabled continuous monitoring, real-time data collection, and intelligent decision-making across a wide range of domains~\cite{jamshed2022challenges}. These sensors, often embedded in wearable devices, home appliances, and environmental systems, provide seamless and scalable data acquisition for applications in healthcare, environmental monitoring, and education. However, their deployment still faces several challenges, including data synchronization, latency, and ensuring reliable communication.

One domain where wireless sensor technology shows particularly promising impact is education, through its application in Multimodal Learning Analytics (MMLA). MMLA is an emerging field that focuses on the collection and analysis of complex and heterogeneous data to better understand how learning occurs, with the ultimate goal of optimizing it. This approach expands beyond traditional log data from learning platforms by incorporating physiological and contextual data, such as biometric signals and observable behaviors, captured in both virtual and physical learning environments~\cite{giannakos2022multimodal}.

Physiological data in MMLA are typically gathered through sensors such as eye trackers, EEG headbands, smartwatches, or webcams, which are used to monitor students as they engage in learning tasks. By integrating signals such as eye movements, brain activity, posture, or facial expressions, researchers aim to construct a global and holistic view of the learning process~\cite{sharma2020multimodal}.

However, the integration of these multimodal data sources poses significant challenges. As noted in~\cite{chango2022review,becerra2023m2lads,becerra2025m2lads}, one of the major obstacles in MMLA research is the technical complexity of synchronizing and fusing data from diverse sensors, which often operate at different temporal resolutions and require specialized handling. These difficulties are exacerbated when multiple devices are used simultaneously in real-world environments, where data loss, latency, and limited access to raw signals are common. Consequently, the development of dedicated data acquisition tools that enable real-time, synchronized communication with wearable sensors becomes essential to the success of MMLA applications.

Furthermore, issues such as the low scalability of sensor-based systems, their limited readiness for deployment in authentic educational settings, and the lack of established ethical guidelines for managing sensitive multimodal data can hinder the integration of MMLA into mainstream educational practice~\cite{yan2022scalability}.

To address these challenges, this article presents Watch-DMLT (an acronym for \textit{Data Monitoring and Logging Tool for Smartwatches enabling real-time synchronization and activity data storage}), a data acquisition application for Fitbit Sense 2. It enables real-time collection, synchronization, and storage of physiological and motion data across multiple devices. Unlike most commercial solutions, Watch-DMLT provides high-frequency access to sensor data such as heart rate, acceleration or orientation, features often restricted in consumer-grade devices. Leveraging wearable capabilities, this work offers a scalable, accessible solution tailored to MMLA research in real learning environments.

In addition, Watch-DMLT has been tested in classroom settings during oral presentation monitoring, where heart rate and motion data were collected from 65 students. To support the validation and multimodal alignment of these signals, a complementary system called ViSeDOPS (\textit{System for Visualisation of Sensor Data from Oral Presentations}) was also developed. ViSeDOPS allows researchers to synchronize and visualize smartwatch data with other modalities, such as video recordings of the session and eye-tracking glasses data facilitating the visual inspection and analysis thought interactive web dashboards.

The data collection process adhered strictly to ethical standards: all participants provided informed consent prior to participation, and data were fully anonymized to protect individual identities. Sensitive information was securely stored following established privacy protocols.

The remainder of this paper is structured as follows. Section \ref{s:related_work} reviews related work on multimodal learning analytics and the use of wearable sensors in educational contexts. Section \ref{s:watch_dmlt} describes the design and technical implementation of Watch-DMLT. Section \ref{s:monitoring} presents a real-world classroom deployment and details the multimodal monitoring setup using Watch-DMLT and ViSeDOPS. In Section \ref{s:conclusions}, we discuss conclusions and outline directions for future work.

\section{Related Work}\label{s:related_work}
\subsection{Sensors in Multimodal Learning Analytics}
The use of sensors in MMLA has gained significant traction due to their potential to capture cognitive, affective, and behavioral aspects of learning that are otherwise difficult to observe. Sensors such as eye-trackers, EEG devices, cameras, and smartwatches are increasingly employed to gather physiological and behavioral data. These tools allow researchers to model higher-order constructs like engagement, motivation, cognitive load, and emotional states in real time \cite{cukurova2020promise}. The fusion of sensor-based multimodal data leads to more accurate predictions of learning outcomes and provides a richer understanding of learning processes. For instance, combining gaze data with EEG, heart rate, and video has shown improved prediction performance compared to unimodal approaches \cite{giannakos2019multimodal}.

Another context where sensors have proven particularly valuable is online learning. MMLA has demonstrated high effectiveness in these environments by leveraging biometric and behavioral data captured through learners’ digital activity, such as webcam recordings, eye-tracking, and physiological signal monitoring \cite{baro2018integration,daza2023edbb}. These sensor-based platforms benefit from recent advances in machine learning and behavioral modeling, enabling a deeper understanding of cognitive and emotional states and fostering more personalized and adaptive learning experiences. In particular, these developments have enabled the creation of large-scale multimodal datasets such as IMPROVE \cite{daza2024improve}, which monitored 120 learners and combines 16 synchronized sensor streams, including EEG, smartwatches, gaze tracking, video, and input dynamics, to study how mobile phone usage affects attention and learning performance in online education.


Smartwatches have become increasingly popular in MMLA due to their ability to unobtrusively capture continuous physiological signals such as heart rate, heart rate variability, and electrodermal activity. These signals are valuable proxies for cognitive load, stress, attention, and emotional arousal during learning tasks \cite{larmuseau2020multimodal}. For instance, heart rate has been used to monitor attention levels, with findings indicating that a decrease in heart rate can serve as a reliable marker of fatigue and reduced alertness \cite{kawamura2021detecting}. In \cite{becerra2024biometrics} analyzed heart rate variations before, during, and after mobile phone usage and found that learners showed significantly higher heart rate values while responding to phone messages. This increase was interpreted as a stress-related physiological reaction, supporting the idea that heart rate data from smartwatches can serve as a meaningful indicator of cognitive and emotional responses. In \cite{becerra2025ai}, heart rate data collected from smartwatches was used to detect smartphone use during online learning. Their findings showed that heart rate alone could predict phone usage with 70\% accuracy, and when combined with other signals in a multimodal model, accuracy improved to 91\%, confirming its value as an indicator of distraction.

\subsection{Multimodal Learning Analytics Dashboards}

Dashboards have been widely used in MMLA to visualize and extract insights from multimodal student data. For instance, the system presented in \cite{becerra2023m2lads,becerra2025m2lads}, M2LADS, integrates and visualizes multimodal data collected during MOOC sessions. M2LADS aggregates biometric signals such as heart rate, visual attention, and brain activity to provide a comprehensive view of learners’ interactions with the course. The system generates dashboards that enable instructors to monitor students’ attention and engagement levels during specific learning activities. Another example is VAAD \cite{navarro2024vaad}, a tool specifically designed to visualize and analyze eye-tracking data gathered during online learning sessions.

Recent research has increasingly emphasized the design of human-centered dashboards in educational contexts, integrating AI and Generative AI to enhance not only functionality but also usability, interpretability, and pedagogical alignment \cite{topali2025designing}. Examples of these type of Dashboards are VizChat \cite{yan2024vizchat} GePeTo \cite{becerra2024generative} or AICoFe \cite{becerra2025aicofe}.

\section{Watch-DMLT}\label{s:watch_dmlt}
Watch-DMLT is an application developed to collect physiological and motion-related signals, including heart rate, acceleration, gyroscope, and orientation data, from individuals wearing a Fitbit Sense 2 smartwatch. While commercial smartwatches are primarily designed for individual consumer use, they often lack the real-time and high-resolution data access capabilities required for research or observational studies.

One of the main limitations of commercial devices is their restricted access to real-time data. Typically, on a weekly or monthly basis, and do not support real-time data retrieval. In particular, Fitbit devices restrict access to raw data from sensors such as the gyroscope, accelerometer, and orientation module, which are essential for detailed activity tracking. Moreover, the heart rate sensor often samples at low frequencies (e.g., one reading every five seconds) and may experience intermittent data loss, limiting its reliability for precise analysis. Furthermore, in scenarios where synchronized data collection from multiple devices is required, such as in collaborative tasks or group studies, commercial solutions generally fall short. Watch-DMLT supports real-time, parallel data acquisition from multiple smartwatches without interference or data overlap.

Watch-DMLT has been developed using the Fitbit Software Development Kit (SDK) \footnote{\url{https://dev.fitbit.com/getting-started/}}, which offers a comprehensive set of tools, libraries, documentation, and sample code for application development on Fitbit devices. Through this SDK, access is granted to various integrated sensors such as the photoplethysmographic heart rate monitor, accelerometer, gyroscope, and orientation sensor.

During monitoring sessions, the application records data from each sensor and stores it in .csv files. Each file contains the date, timestamp, and the corresponding sensor-specific measurement for each entry. Specifically, the recorded measurements include: beats per minute for heart rate; angular velocity along three orthogonal axes (X, Y, Z) from the gyroscope; linear acceleration along three orthogonal axes (X, Y, Z) from the accelerometer; and quaternion components, scalar part $w$ and vector parts $x,y,z$, representing device orientation.

Watch-DMLT comprises three main functional modules:

\begin{itemize}
    \item \textbf{Data Collection:} Sensor data are captured using the Fitbit Sense 2 smartwatch sensors. This process is managed through on-screen visual components and concludes with the transmission of the data to a paired mobile device (see Subsection~\ref{SEC:RELOJ}).

    \item \textbf{Data Transmission:} This module runs on a mobile device (in this case, a Redmi 9C) paired with the smartwatch. It receives the transmitted files and uploads them to a server using an HTTP POST request (see Subsection~\ref{SEC:MOVIL}).

    \item \textbf{Data Processing:} On the server side, received data is unified and stored for subsequent analysis (see Subsection~\ref{SEC:SERVIDOR}).
\end{itemize}

\begin{figure}[t]
\centering
\includegraphics[width=\textwidth]{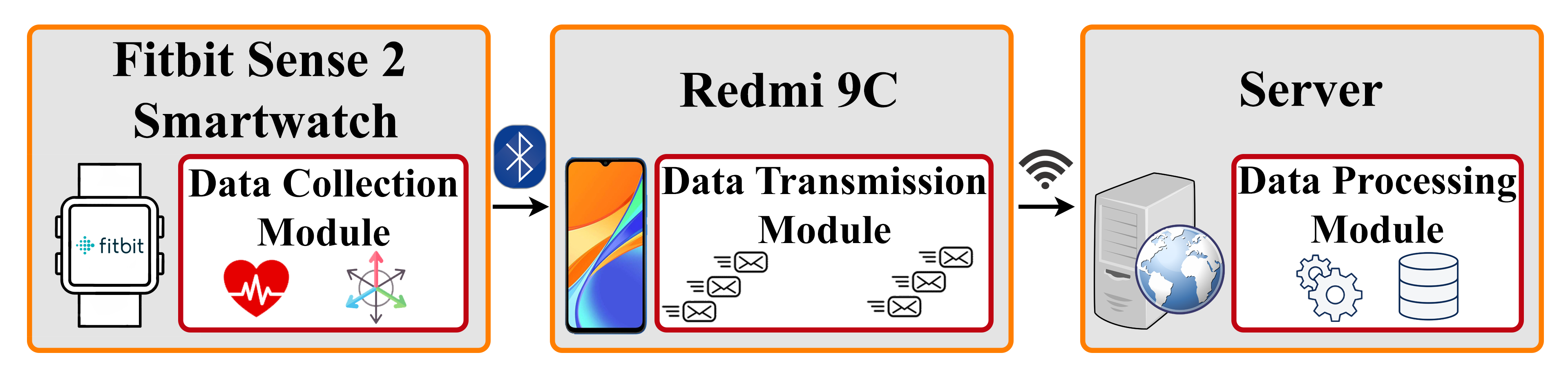}
\caption{Data communication architecture between smartwatch,  phone, and server.}
\label{FIG:Comunicacion}
\end{figure}

Communication between the data collection and transmission modules is handled via a message queue that ensures orderly and reliable transfer of generated files. Bluetooth connectivity between the smartwatch and mobile device is essential for this communication. Likewise, HTTPS is used to securely transmit data between the mobile device and the server (Figure~\ref{FIG:Comunicacion}).

\subsection{Data Collection Module}
\label{SEC:RELOJ}

The data collection module runs directly on the Fitbit Sense 2 smartwatch and interfaces with the device’s onboard sensors to acquire physiological and motion-related signals. It also manages the smartwatch’s user interface, allowing manual control of recording sessions via a start/stop button, and includes ``+'' and ``–'' buttons to assign a unique device identifier, crucial for distinguishing data in multi-device setups.

Two main limitations were addressed during development, both due to hardware constraints. The limited RAM prevents prolonged data buffering, so the application periodically writes data to local files to ensure stability. Similarly, the smartwatch's restricted storage capacity required implementing periodic data transmission to an external module for server upload.

Determining the transmission frequency was critical: intervals shorter than 30 seconds risked overloading the communication queue, while those longer than two minutes led to memory overflow. Through testing, a one-minute interval was found to offer a reliable balance between performance and data integrity.

\subsection{Data Transmission Module}
\label{SEC:MOVIL}

The data transmission module runs on the Fitbit mobile application installed on the smartphone paired with the smartwatch. It receives the data files generated by the collection module through a message queue and forwards them to the server using secure HTTPS requests.

This process relies on proper synchronization between the smartwatch and the smartphone. Successful data transfer requires that both devices are connected to a Wi-Fi network and synchronized before and after each monitoring session. Since biometric data is stored locally on the smartwatch, failure to synchronize promptly can result in transmission delays or data loss if the connection is not re-established in time.

\subsection{Data Processing Module}
\label{SEC:SERVIDOR}

The server-side data processing module is responsible for receiving, organizing, and storing the files transmitted from the mobile device. Communication between the transmission module and the server is established via the HTTPS protocol, in compliance with the security requirements enforced by the Fitbit Software Development Kit (SDK).

To overcome this limitation, the server employs Ngrok\footnote{\url{https://ngrok.com}}, a tunneling service that enables secure exposure of a local server to the public Internet. Ngrok provides a publicly accessible HTTPS endpoint that acts as a secure bridge, allowing the mobile application to transmit data to the server.

Once the data files arrive at the server, they are processed by a unification mechanism that consolidates all files associated with the same smartwatch identifier. This process ensures that data collected from each individual device is grouped and stored in a single, structured file corresponding to the monitoring session. Specifically, four distinct files are generated per participant, each representing a different sensor stream. This automated consolidation facilitates immediate availability of the data for downstream analysis, eliminating the need for manual pre-processing or merging tasks.

\section{Multimodal Monitoring with Watch-DMLT}\label{s:monitoring}

To evaluate the capabilities of Watch-DMLT in real-world educational settings, we conducted a case study involving 65 students delivering oral presentations at Universidad Autónoma de Madrid (UAM)~\cite{becerra2025mosaicf}. During these sessions, multimodal data were collected in real time using a set of devices and later processed using ViSeDOPS, a dashboard-based visualization system specifically developed for this purpose. ViSeDOPS integrates, cleans and synchronizes fully anonymized data from various sources (Figure \ref{fig:visedops_diagrama}), including:

\begin{figure}[t]
\centering
\includegraphics[width=\textwidth]{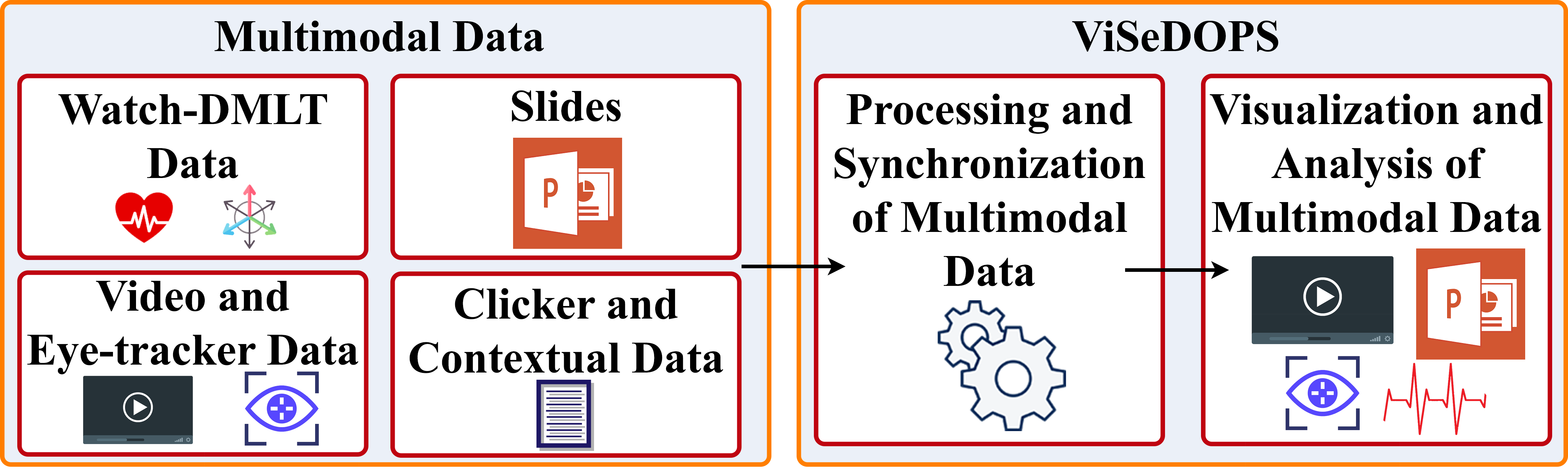}
\caption{Overview of the ViSeDOPS system architecture.}
\label{fig:visedops_diagrama}
\end{figure}

\begin{itemize}
    \item \textbf{Logitech C920 PRO HD Webcam:} A front-facing RGB camera was used to record the presenter’s delivery, capturing both video and audio.
    
    \item \textbf{Fitbit Sense 2 Smartwatches:} Up to 16 smartwatches operated in parallel using Watch-DMLT to collect heart rate and motion data from both presenters and audience members.
    
    \item \textbf{Clicker, Mouse, and Keyboard:} User interactions with presentation tools, such as advancing slides, were automatically logged to contextualize physiological responses and timing.
    
    \item \textbf{Tobii Pro Glasses 3:} An eye-tracking device worn by an audience member captured gaze behavior, including fixations and saccades, offering an external viewpoint of attention during the session.
    
    \item \textbf{Contextual Data Annotations:} A research assistant labelled events in real time (e.g., nervous gestures, eye contact, or reading from slides), enriching the dataset with qualitative behavioral markers.
\end{itemize}

Several key challenges were encountered while using Watch-DMLT in this real classroom environment. As the number of devices increased, managing device assignment, and synchronization became progressively more complex, requiring careful coordination and manual tracking. Additionally, the paired smartphone needed to remain in close proximity to each monitored student. When the phone was too far from the smartwatch, the risk of data loss and message queue overflow increased due to delays in file transmission.

Once the data were collected and synchronized, ViSeDOPS generated an individualized dashboard for each student, enabling in-depth multimodal analysis. These dashboards are implemented using the \textit{Plotly Dash} framework\footnote{\url{https://dash.plotly.com/}}, a Python library for building interactive web application. The dashboards include:
\begin{itemize}
    \item \textbf{Heart Rate Graphs:} Physiological responses of presenters and audience members are visualized through dynamic line graphs. Data can be segmented by slide or annotated events, allowing for targeted analysis of stress or arousal in response to specific moments. Summary statistics (e.g., maximum, mean) and paired statistical comparisons are provided to highlight variations.

\item \textbf{Synchronized Video Playback:} The dashboard incorporates video recordings of the presentation and the eye-tracker view. Selecting a timestamp in the heart rate graph automatically aligns and plays the corresponding video frame, facilitating contextual inspection of physiological changes.

\item \textbf{Presentation Slides:} All slides used during the session are displayed within the dashboard. Automated analysis extracts design features such as font type, size, and visible numbering. This allows researchers to examine how visual structure may influence cognitive or emotional responses.
\end{itemize}

\begin{figure}[t]
\centering
\includegraphics[width=0.8\textwidth]{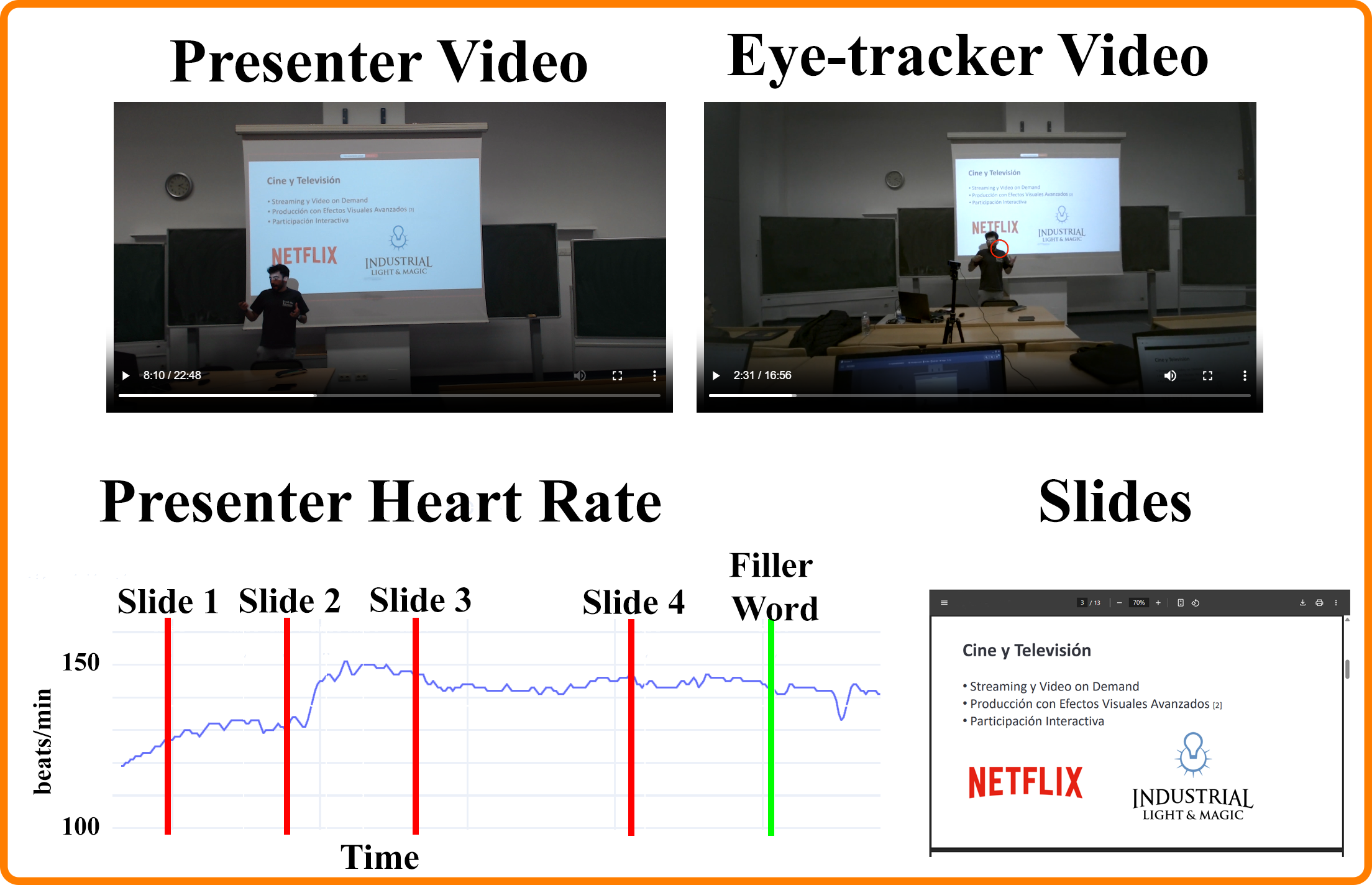}
\caption{Some visualizations extracted from ViSeDOPS.}
\label{fig:visedops}
\end{figure}

An example of these visualizations is shown in Figure~\ref{fig:visedops}. Together, Watch-DMLT and ViSeDOPS provide a scalable and extensible infrastructure for collecting and interpreting multimodal data in real classroom environments, offering valuable insights into student behavior and presentation performance.

\section{Conclusions and Future Work}\label{s:conclusions}

This work presents the design, implementation, and validation of two complementary tools, Watch-DMLT and ViSeDOPS. Watch-DMLT is an application for Fitbit Sense 2 smartwatches that enables scalable, real-time acquisition of physiological and motion data from multiple smartwatches, overcoming the typical limitations of commercial wearables. ViSeDOPS complements this functionality by providing a flexible and interactive dashboard that synchronizes multimodal data sources, including heart rate, motion, eye-tracking, video, and contextual annotations, to support fine-grained analysis of student performance.

Through a classroom deployment involving 65 students and up to 16 synchronized devices, we demonstrate the feasibility of using Watch-DMLT in real educational scenarios. Furthermore, the dashboards generated by ViSeDOPS allow professors and researchers to examine physiological responses in context, identify patterns of stress and engagement.

For future work we plan to extend the use of Watch-DMLT and ViSeDOPS to other educational contexts, such as online learning and collaborative tasks. Additionally, we aim to enhance the analytical capabilities of ViSeDOPS by incorporating predictive modeling and conducting more in-depth analyses.

\section*{Acknowledgements}
Support by projects: Cátedra ENIA UAM-VERIDAS en IA Responsable (Next- \\GenerationEU PRTR TSI-100927-2023-2),
HumanCAIC (TED2021-131787B-I00ICINN), and SNOLA (RED2022-134284-T).

%
%
\bibliographystyle{splncs04}
\bibliography{biblio}

@book{giannakos2022multimodal,
  title={{The Multimodal Learning Analytics Handbook}},
  author={Giannakos, Michail and Spikol, Daniel and Di Mitri, Daniele and Sharma, Kshitij and Ochoa, Xavier and Hammad, Rawad},
  year={2022},
  publisher={Springer}
}

@inproceedings{becerra2025aicofe,
  title={{Enhancing the Professional Development of Engineering Students Through an AI-Based Collaborative Feedback System}},
  author={Becerra, Alvaro and Cobos, Ruth},
  booktitle={{2025 IEEE Global Engineering Education Conference (EDUCON)}},
  pages={1--9},
  year={2025},
  organization={IEEE}
}

@inproceedings{becerra2024generative,
  title={{A Generative AI-Based Personalized Guidance Tool for Enhancing the Feedback to MOOC Learners}},
  author={Becerra, Alvaro and Mohseni, Zeynab and Sanz, Javier and Cobos, Ruth},
  booktitle={{2024 IEEE Global Engineering Education Conference (EDUCON)}},
  pages={1--8},
  year={2024},
  organization={IEEE}
}

@inproceedings{daza2023edbb,
  title={{edBB-Demo: Biometrics and Behavior Analysis for Online Educational Platforms}},
  author={Daza, Roberto and Morales, Aythami and Tolosana, Ruben and Gomez, Luis F and Fierrez, Julian and Ortega-Garcia, Javier},
  booktitle={{Proceedings of the AAAI Conference on Artificial Intelligence}},
  volume={37},
  number={13},
  pages={16422--16424},
  year={2023}
}

@article{baro2018integration,
  title={{Integration of an Adaptive Trust-Based E-Assessment System into Virtual Learning Environments—The TeSLA Project Experience}},
  author={Bar{\'o}-Sol{\'e}, Xavier and Guerrero-Roldan, Ana E and Prieto-Bl{\'a}zquez, Josep and Rozeva, Anna and Marinov, Orlin and Kiennert, Christophe and Rocher, Pierre-Olivier and Garcia-Alfaro, Joaquin},
  journal={Internet Technology Letters},
  volume={1},
  number={4},
  pages={e56},
  year={2018},
  publisher={Wiley Online Library}
}

@article{daza2024improve,
  title={{IMPROVE: Impact of Mobile Phones on Remote Online Virtual Education}},
  author={Daza, Roberto and Becerra, Alvaro and Cobos, Ruth and Fierrez, Julian and Morales, Aythami},
  journal={arXiv preprint arXiv:2412.14195},
  year={2024}
}

@article{becerra2025m2lads,
  title={{M2LADS Demo: A System for Generating Multimodal Learning Analytics Dashboards}},
  author={Becerra, Alvaro and Daza, Roberto and Cobos, Ruth and Morales, Aythami and Fierrez, Julian},
  journal={arXiv preprint arXiv:2502.15363},
  year={2025}
}

@inproceedings{becerra2023m2lads,
  title={{M2LADS: A System for Generating Multimodal Learning Analytics Dashboards}},
  author={Becerra, Alvaro and Daza, Roberto and Cobos, Ruth and Morales, Aythami and Cukurova, Mutlu and Fierrez, Julian},
  booktitle={{2023 IEEE 47th Annual Computers, Software, and Applications Conference (COMPSAC)}},
  pages={1564--1569},
  year={2023},
  organization={IEEE}
}

@inproceedings{navarro2024vaad,
  title={{VAAD: Visual Attention Analysis Dashboard Applied to E-Learning}},
  author={Navarro, Miriam and Becerra, Alvaro and Daza, Roberto and Cobos, Ruth and Morales, Aythami and Fierrez, Julian},
  booktitle={{2024 International Symposium on Computers in Education (SIIE)}},
  pages={1--6},
  year={2024},
  organization={IEEE}
}

@inproceedings{becerra2024biometrics,
  title={{Biometrics and Behavioral Modelling for Detecting Distractions in Online Learning}},
  author={Becerra, Alvaro and Irigoyen, Javier and Daza, Roberto and Cobos, Ruth and Morales, Aythami and Fierrez, Julian and Cukurova, Mutlu},
  booktitle={{Proc. Simposio Internacional de Inform\'{a}tica Educativa (SIIE), VII Congreso Espa\~{n}ol de Inform\'{a}tica}},
  year={2024}
}

@article{sharma2020multimodal,
  title={Multimodal Data Capabilities for Learning: What Can Multimodal Data Tell Us about Learning?},
  author={Sharma, Kshitij and Giannakos, Michail},
  journal={British Journal of Educational Technology},
  volume={51},
  number={5},
  pages={1450--1484},
  year={2020},
  publisher={Wiley Online Library}
}

@article{chango2022review,
  title={{A Review on Data Fusion in Multimodal Learning Analytics and Educational Data Mining}},
  author={Chango, Wilson and Lara, Juan A and Cerezo, Rebeca and Romero, Crist{\'o}bal},
  journal={Wiley Interdisciplinary Reviews: Data Mining and Knowledge Discovery},
  volume={12},
  number={4},
  pages={e1458},
  year={2022},
  publisher={Wiley Online Library}
}

@inproceedings{yan2022scalability,
  title={{Scalability, Sustainability, and Ethicality of Multimodal Learning Analytics}},
  author={Yan, Lixiang and Zhao, Linxuan and Gasevic, Dragan and Martinez-Maldonado, Roberto},
  booktitle={LAK22: 12th international learning analytics and knowledge conference},
  pages={13--23},
  year={2022}
}

@article{jamshed2022challenges,
  title={{Challenges, Applications, and Future of Wireless Sensors in Internet of Things: A Review}},
  author={Jamshed, Muhammad Ali and Ali, Kamran and Abbasi, Qammer H and Imran, Muhammad Ali and Ur-Rehman, Masood},
  journal={IEEE Sensors Journal},
  volume={22},
  number={6},
  pages={5482--5494},
  year={2022},
  publisher={IEEE}
}

@article{cukurova2020promise,
  title={The Promise and Challenges of Multimodal Learning Analytics},
  author={Cukurova, Mutlu and Giannakos, Michail and Martinez-Maldonado, Roberto},
  journal={British Journal of Educational Technology},
  volume={51},
  number={5},
  pages={1441--1449},
  year={2020},
  publisher={Wiley}
}

@article{giannakos2019multimodal,
  title={Multimodal Data as a Means to Understand the Learning Experience},
  author={Giannakos, Michail N and Sharma, Kshitij and Pappas, Ilias O and Kostakos, Vassilis and Velloso, Eduardo},
  journal={International Journal of Information Management},
  volume={48},
  pages={108--119},
  year={2019},
  publisher={Elsevier}
}

@article{larmuseau2020multimodal,
  title={{Multimodal Learning Analytics to Investigate Cognitive Load during Online Problem Solving}},
  author={Larmuseau, Charlotte and Cornelis, Jan and Lancieri, Luigi and Desmet, Piet and Depaepe, Fien},
  journal={British Journal of Educational Technology},
  volume={51},
  number={5},
  pages={1548--1562},
  year={2020},
  publisher={Wiley Online Library}
}

@article{kawamura2021detecting,
  title={Detecting Drowsy Learners at the Wheel of e-learning Platforms with Multimodal Learning Analytics},
  author={Kawamura, Ryosuke and Shirai, Shizuka and Takemura, Noriko and Alizadeh, Mehrasa and Cukurova, Mutlu and Takemura, Haruo and Nagahara, Hajime},
  journal={IEEE Access},
  volume={9},
  pages={115165--115174},
  year={2021},
  publisher={IEEE}
}

@article{topali2025designing,
  title={Designing Human-centered Learning Analytics and Artificial Intelligence in Education Solutions: a Systematic Literature Review},
  author={Topali, Paraskevi and Ortega-Arranz, Alejandro and Rodr{\'\i}guez-Triana, Mar{\'\i}a Jes{\'u}s and Er, Erkan and Khalil, Mohammad and Ak{\c{c}}ap{\i}nar, G{\"o}khan},
  journal={Behaviour \& Information Technology},
  volume={44},
  number={5},
  pages={1071--1098},
  year={2025},
  publisher={Taylor \& Francis}
}

@article{becerra2025mosaicf,
  title={{MOSAIC-F: A Framework for Enhancing Students' Oral Presentation Skills through Personalized Feedback}},
  author={Becerra, Alvaro and Andres, Daniel and Villegas, Pablo and Daza, Roberto and Cobos, Ruth},
  journal={arXiv preprint arXiv:2506.08634},
  year={2025}
}

@inproceedings{yan2024vizchat,
  title={{VizChat: Enhancing Learning Analytics Dashboards with Contextualised Explanations using Multimodal Generative AI Chatbots}},
  author={Yan, Lixiang and Zhao, Linxuan and Echeverria, Vanessa and Jin, Yueqiao and Alfredo, Riordan and Li, Xinyu and Ga{\v{s}}evi’c, Dragan and Martinez-Maldonado, Roberto},
  booktitle={International Conference on Artificial Intelligence in Education},
  pages={180--193},
  year={2024},
  organization={Springer}
}

@article{becerra2025ai,
  title={{AI-based Multimodal Biometrics for Detecting Smartphone Distractions: Application to Online Learning}},
  author={Becerra, Alvaro and Daza, Roberto and Cobos, Ruth and Morales, Aythami and Cukurova, Mutlu and Fierrez, Julian},
  journal={arXiv preprint arXiv:2506.17364},
  year={2025}
}
\end{document}